\documentclass{iau}

\usepackage{aas_macros}
\usepackage{amsmath}
\usepackage{graphicx}
\usepackage{multirow}
\usepackage{orcidlink}
\usepackage[inline]{enumitem}

\newcommand{\msun}{\ensuremath{\mathrm{M}_\odot}}

\newcommand{\nbo}{\texttt{NBODY6++GPU}}

\newcommand{\nbody}{\textit{N}-body}
\newcommand{\done}{\texttt{DRAGON-I}}
\newcommand{\dtwo}{\texttt{DRAGON-II}}
\newcommand{\dthree}{\texttt{DRAGON-III}}

\newcommand{\fref}[1]{Fig.~\ref{#1}}
\newcommand{\tref}[1]{Table~\ref{#1}}

\newcommand{\ff}[1]{\textcolor{black}{#1}}
\newcommand{\rainer}[1]{\textcolor{black}{#1}}

\begin{document}

\lefttitle{DRAGON-III simulation: modelling million-body globular and nuclear star clusters}
\righttitle{DRAGON-III simulation: modelling million-body globular and nuclear star clusters}

\jnlPage{1}{7}
\jnlDoiYr{2021}
\doival{10.1017/xxxxx}

\aopheadtitle{Proceedings IAU Symposium}
\editors{H. M. Lee \&  R. Spurzem , eds.}

\title{\texttt{DRAGON-III} simulations: modelling million-body globular and nuclear star clusters over cosmic time}

\author{Kai Wu$^{1}$\footnote{\email{kaiwu.astro@gmail.com}}, Philip Cho$^{1}$, Rainer Spurzem$^{1,2,3}$\footnote{\email{spurzem@ari.uni-heidelberg.de}}, Long Wang$^{4}$, \\ Francesco Flammini Dotti$^{5,6}$, Vahid Amiri$^{1}$}
\affiliation{$^{1}$Astronomisches Rechen-Institut, Zentrum f\"{u}r Astronomie der Universit\"{a}t  Heidelberg, M\"onchhofstr. 12-14, D-69120 Heidelberg, Germany}
\affiliation{$^{2}$National Astronomical Observatories, Chinese Academy of Sciences, 20A Datun Rd., Chaoyang District, 100101, Beijing, China}
\affiliation{$^{3}$Kavli Institute for Astronomy and Astrophysics, Peking University, Yiheyuan Lu 5, Haidian Qu, 100871, Beijing, China}
\affiliation{$^{4}$School of Physics and Astronomy, Sun Yat-sen University (Zhuhai Campus), Zhuhai 519082, China}
\affiliation{$^{5}$Department of Physics, New York University Abu Dhabi, PO Box 129188 Abu Dhabi, UAE}
\affiliation{$^{6}$Center for Astrophysics and Space Science (CASS), New York University Abu Dhabi, PO Box 129188, Abu Dhabi, UAE}

\begin{abstract}
As a continuation of \texttt{DRAGON-II}, we present the \texttt{DRAGON-III} project, which focuses on the simulations of million-body globular clusters and nuclear clusters over 10~Gyr. We report on its preliminary results on globular clusters.
The first 100~Myr of the simulations have produced 41 pulsars, 191 X-ray binaries, 17 gravitational wave sources, and one black hole-black hole merger due to the loss of orbital energy in the form of gravitational wave emission. The inclusion of initial soft binaries brings surprisingly interesting results, including one IMBH in a binary black hole, and compact object binaries resembling the Gaia-BH1 and the wide black hole-giant binary reported in \cite{wang2024massgapbh}.
\end{abstract}


\maketitle

\section{Introduction}

\subsection{Multi-messenger astrophysics of globular and nuclear star clusters}

Globular clusters (GCs), abundant in galactic disks and spheroids, serve as ideal laboratories for studying stellar evolution alongside Newtonian and relativistic dynamics. Nuclear star clusters (NSCs), which often host supermassive black holes in massive elliptical galaxies, generate tidal disruption events as stars or compact objects interact with the central black hole. 

From a comprehensive set of computer models, multi-messenger observational signatures of GCs and NSCs can be deduced, including 
\begin{enumerate*} 
    \item star-by-star kinematic information for nearby star clusters (Gaia, LSST, JWST),
    \item numbers of X-ray binaries in local globular clusters and in our own Galactic Center,
    \item gravitational wave (GW) signals (and possible electromagnetic counterparts) of merging binary black holes (BHs) and neutron stars (NSs) in a local universe (LIGO/Virgo/KAGRA and future Einstein Telescope \citealt{branchesi2023ET} Cosmic Explorer \citealt{ricci2015alma}), and
    \item gravitational inspiral events produced by objects falling into BHs to be discovered years before their coalescence (LISA \citealt{amaro-seoane2017lisa}, TianQin \citealt{luo2016tianqin}, and Taiji \citealt{chen2021taiji-lisa,ruan2021the-lisa-taiji}).
\end{enumerate*}

\subsection{The \texttt{DRAGON} simulation family}

The \texttt{DRAGON} simulations \ff{aim} to model the largest star clusters to the most realistic degree, \ff{using} accurate \nbody{} methods with up-to-date treatment for stellar \ff{evolution recipes}. 

\done{} \citep{wang2016dragon} is the first \nbody{} simulation to simulate million-body GCs for Giga-years 
while \dtwo{} \citep{dragon2-2,dragon2-1,dragon2-3} significantly improves stellar evolution models involving compact objects, and predicts gravitational wave signals from compact object mergers that would be measured by LIGO/Virgo/KAGRA. 

The aim of \dthree{} is to get step by step closer to realistic particle numbers for massive globular and nuclear star clusters. Furthermore, the inclusion of a novel feature, a large number of initially wide binaries, significantly changes the entire binary population during the evolution. Moreover, we follow all escaping stars to investigate how they build tidal tails with a more recent galaxy model.

\section{Numerical methods}

\subsection{Code}

\texttt{\nbo{}} \citep{NBODY6++GPU} is one of the state-of-the-art codes capable of modelling star clusters to the most realistic degree possible. It has a long history as an offspring of Sverre's \nbody{} code \citep{aarseth1999industry}, and is being continuously maintained by the Silkroad Team \footnote{\url{https://www.astro-silkroad.eu}} and made publicly available \footnote{\url{https://github.com/nbody6ppgpu}}, with a major update for stellar evolution by \citet{NBODY6++GPUNewSSE}. We refer readers to the review article \citet{spurzem2023rev} for its physics, algorithms, features, and achievements. Beyond the descriptions therein, we recently ported the \ff{G}alactic potential \texttt{MWPotential2014} from \texttt{Galpy} \citep{bovy2015galpy}, for better comparison with results from other codes. In recent years, the \texttt{MWPotential2014} has been adopted by other star cluster codes (e.g., \texttt{MOCCA} by \citealt{giersz1998monte}; \texttt{PeTar} by \citealt{wang2020petar}).
    We have ported the \texttt{CUDA} code in \nbo{} to \texttt{HIP}\footnote{\url{https://rocm.docs.amd.com/projects/HIP/en/docs-develop/what_is_hip.html}} software stacks, enabling our program to run large-scale simulations on next-generation non-CUDA architectures, including DCU-accelerated computing clusters from China and AMD-GPU-based LUMI supercomputer in Finland.

\texttt{PeTar}\footnote{\url{https://github.com/lwang-astro/PeTar}} is one of the most popular codes for star cluster \nbody{} simulations. It was first developed by \citet{wang2020petar}, and then enriched with functions such as \texttt{BSE} \citep{wang2021petarbse}, \texttt{BSEEMP} \citep{wang2022petarbseemp}, and \texttt{MWPotential2014} \citep{wang2021petargalpy}. We refer interested readers to the article of \citet{wang2025iauproceeding} in the same proceeding volume. 

\subsection{Initial conditions}

\tref{d3ic} shows the initial conditions of the \rainer{ongoing and} planned \dthree{} simulations. We follow the model with 1 million stars in \dtwo{} for unlisted initial parameters, and refer readers to \citet[][Section 2.3]{dragon2-1} for them. Inspired by \citet{offner2023soft-binary}, we explore three different fractions of initial soft (wide) binaries. We do not remove stars from simulations when they are far enough from the cluster centre and usually considered as \textit{escaping}. Instead, we keep evolving them in the galactic tidal field. This allows us to investigate possible stellar streams formed by massive clusters. 

\begin{table}
 \centering
 \caption{Initial conditions for \dthree{} simulations}\label{d3ic}
   {\tablefont
    \begin{tabular}{rcc}
      \midrule\midrule
       & \dthree{} GC & \dthree{} NSC \\
       &  & with accreting CMBH \\
      \midrule
      Number of stars & 1M, 4M, 10M & 1M \\
      Integration Time & 1 -- 10 Gyr & $\sim$ Gyr \\
      Orbit in the Galaxy & $a = 10$~kpc; $e=0.5$; $inc = 60^\circ$ & In the galaxy centre \\
      \midrule
      Tidal field & \multicolumn{2}{c}{Galpy \texttt{MWPotential2014} } \\
      Binary & \multicolumn{2}{c}{hard: 5\% \& soft: 0\% / 20\% / 60\%} \\
      Stellar evolution packages & \multicolumn{2}{c}{\citet{NBODY6++GPUNewSSE,banerjee2020bse-versus,banerjee2021stellar-mass}} \\
      IMF \& profile & \multicolumn{2}{c}{\citet{kroupa2001a-IMF} 0.08 -- 150 \msun{} \& \citet{kingprofile}} \\
      \midrule\midrule
    \end{tabular}
    }
\tabnote{\textit{Notes}: [1] The accreting central supermassive black hole is added in NSC simulations. [2] The orbit of the GC is taken typical observational values from \citet{vasilievBaumgardt2021gaia,bajkova2021orbits} [3] In the initial setup, \textit{hard binaries} means those with semi-major axis of $0.1~\mathrm{au} \leq a \leq 50~\mathrm{au}$, while \textit{soft binary} means $50~\mathrm{au} \le a \leq 5000~\mathrm{au}$. }
\end{table}

\section{Selected preliminary results}

\begin{figure}
  \centering
  \includegraphics[width=0.84\columnwidth]{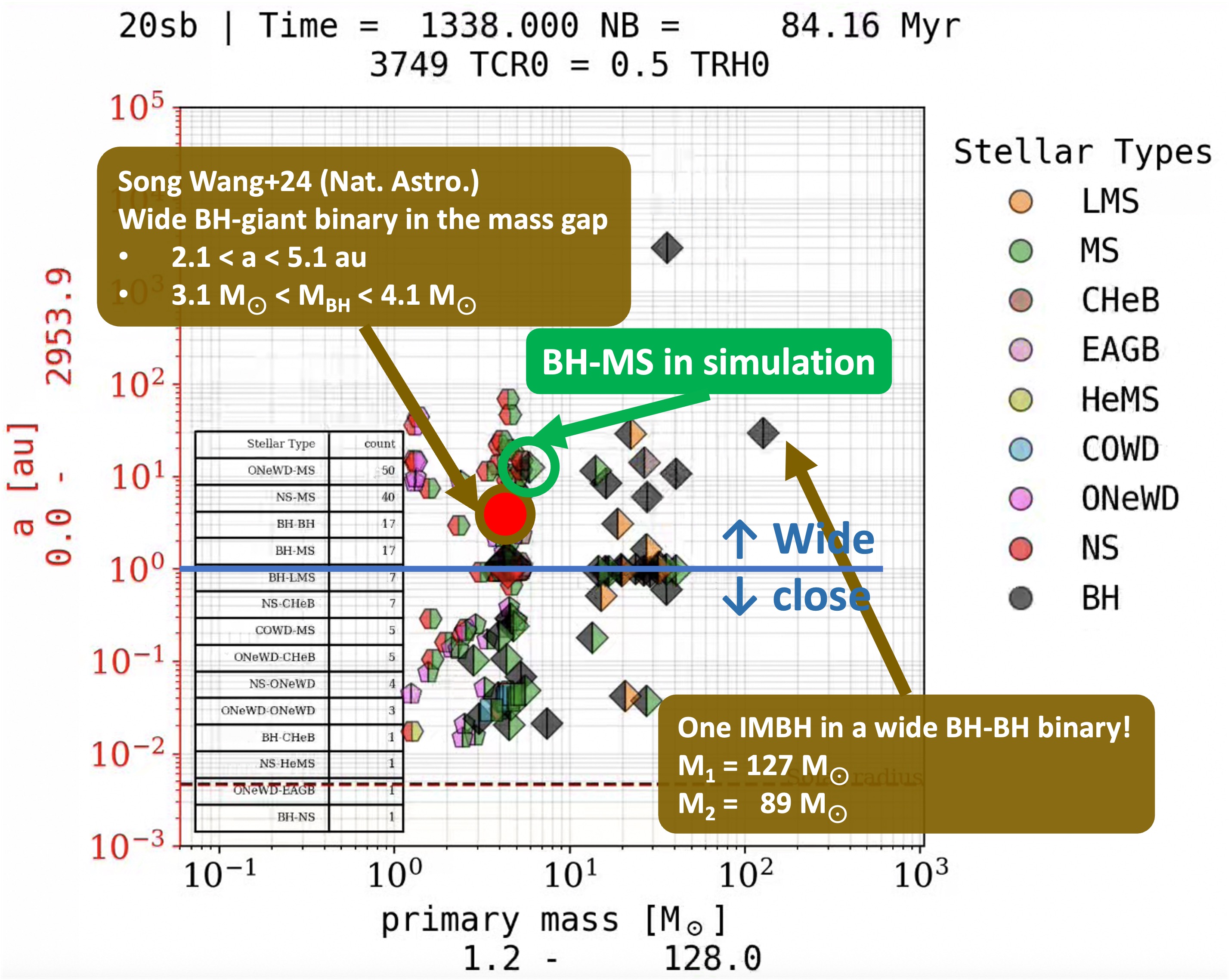}
   \caption{Semi-major axis as a function of primary mass for binaries with at least one compact object component, in the simulation with 20\% initial soft binary after evolving for 84.16~Myr, which equals 3749 initial crossing time, and 0.5 initial half-mass relaxation time. In the bottom left corner is the count of binaries by stellar types of their components. Meaning of the stellar type abbreviation in this table and in the legend can be found in \citet{hurley2000-sse}. Meaning of marker style: square - black hole as primary star, octagon - neutron star as primary star, pentagon - white dwarf as primary star. 
}
  \label{fig.a-m1-compact}
\end{figure}

\fref{fig.a-m1-compact} shows binari\ff{es} related to compact objects in early time of one simulation with \nbo{}. Many interesting 
objects appear after evolving for 84.16~Myr, especially:
\begin{itemize}
    \item An IMBH is present in a wide BH--BH binary, with primary mass of 127~\msun{} and secondary mass of 89~\msun{}. This binary is escaping from the star cluster, and will likely be dynamically unbound to the cluster.
    \item An BH--MS binary ($a=1.29$~au, $M_\mathrm{BH}=5.82$~\msun{}) is located next to the recently observed BH--giant wide binary in the mass gap \citep{wang2024massgapbh} ($2.1 < a/\mathrm{au} < 5.1$, $3.1 < M_\mathrm{BH}/\mathrm{\msun{}} < 4.1$). With further evolution, we will see whether this simulated binary will resemble the observed binary.
\end{itemize}

\begin{table}
  \centering
  \caption{Binaries with at least one compact object component, categorized by stellar types of stellar components, in the simulation with 20\% initial soft binary after evolving for 106.8~Myr. Due to page limit, we refer readers to \citet[Section 4]{hurley2000-sse} for the meaning of abbreviations. }
  \label{tab:compact_obj_binary_100myr}
  {\tablefont
  \begin{tabular}{|r|c|c|c|c|c|c|}
    \hline
    Stellar Type & ONeWD--MS & NS--MS & COWD--MS & BH--MS & BH--BH & ONeWD--CHeB  \\
    \hline
    count & 46 & 26 & 26 & 16 & 16 & 14 \\
    \hline
    Stellar Type & BH--LMS & ONeWD--ONeWD & NS--ONeWD & NS--CHeB & ONeWD--HEMS & COWD--LMS  \\
    \hline
    count & 9 & 8 & 7 & 6 & 2 & 2 \\
    \hline
    Stellar Type & ONeWD--LMS & COWD--CHeB & NS--HG & BH--CHeB & BH--NS & NS--HEMS \\
    \hline
    count & 2 & 2 & 1 & 1 & 1 & 1 \\
    \hline
  \end{tabular}
  }
\end{table}

As shown in \tref{tab:compact_obj_binary_100myr}, after evolving the cluster for 100~Myr, there are 41 pulsar sources, 191 X-ray binaries, 17 gravitational wave sources, and one black hole-black hole merger due to losing orbital energy in the form of gravitational wave emission. These results show that young globular clusters produce numerous observational candidates for multi-messenger astrophysics. Some of these binaries will escape from the cluster and scatter into the galactic field. Their distribution and subsequent evolution will be continuously tracked in the running \dthree{} simulations. 

The inclusion of initial soft binaries produced not only more wide binaries, but also more close binaries \rainer{with unexpected properties as discussed above}. An upcoming article \citep{wu2026d3p1} will highlight how different initial soft binaries affect the simulation. The \dthree{} simulations are ongoing. The first \dthree{} data-release will be publicly available soon.

\bibliographystyle{iaulikekai}
\bibliography{used} 

\end{document}